\documentclass[12pt]{article}
\usepackage{textcomp}

\usepackage[ngerman,english]{babel}

\usepackage{makeidx}
\usepackage{booktabs}
\usepackage{makeidx}

\usepackage{algorithm}
\usepackage{algorithmic}
\usepackage{listings}

\usepackage{comment}
\usepackage{amsmath}
\usepackage{amsfonts}
\usepackage{amssymb}
\usepackage{amsthm}
\usepackage{epsf}
\usepackage{psfrag}
\usepackage{graphicx}
\usepackage[subfigure]{caption2}
\usepackage{subfigure}
\usepackage{palatino}
\usepackage{rotating}
\usepackage{wrapfig}

\usepackage{ziffer}

\usepackage{ngerman}
\usepackage{t1enc}
\usepackage[ansinew]{inputenc}

\usepackage{hyperref}

\newcommand{\R}{\mathbb{R}}

\usepackage{color}
\definecolor{hellgrau}{gray}{0.96}
\definecolor{darkblue}{rgb}{0,0,.7}
\definecolor{darkred}{rgb}{.6,0,0}
\definecolor{darkgreen}{rgb}{.30,.69,.10}
\definecolor{orange}{rgb}{0.8,0.3,0.3}

\title{COVID-19 propagation by diffusion - a two-dimensional approach for Germany}
\author{G\"unter B\"arwolff\footnote{mailto:baerwolf@math.tu-berlin.de}\\
Technische Universit\"at Berlin}

\date{}

\begin{document}

\maketitle

\begin{abstract}

Diffusion comes anytime and everywhere. If there is a gradient or a potential difference of a quantity a diffusion process happens and this ends if an equilibrium
is reached only. The concentration of a species maybe such quantity, or the voltage.
An electric currant will be driven by a voltage difference for example.

In this COVID-19 pandemic one observes both regions with low incidence and other ones
with high incidence. The local different people density could be a reason for that.
In populous areas like big cities or congested urban areas higher COVID-19
incidences could be observed than in rural regions.

The aim of this paper consists in the application of a diffusion
concept to describe one possible issue of the the COVID-19 propagation.

This will be discussed for the German situation based on the
quite different incidence data for the different federal states
of Germany.

With this ansatz some phenomenoms of the actual development of the pandemic could be
confirmed. The model gives a possibility to investigate certain scenarios like
border-crossings or local spreading events and their influence on the COVID-19
propagation as well.

\end{abstract}

\section{Introduction and the mathematical model}
The mathematical modeling of COVID-19 with SIR-type models (\cite{kak}, \cite{gb}) leads to
averaged results and does not take unequal peopling or populousness into account.
But it is well known, that these issues play an important role in the
local pandemic evolution.
With the consideration of local-dependent density of people and a diffusion
model we try to resolve the COVID-19 propagation in a finer manner.

What is a good choice of a quantity to describe the COVID-19 spread? The WHO
and national health institutions measure the COVID-19 spread with the
seven-days-incidence (sometimes also the fourteen-days incidence) 
of infected people per 100000 inhabitants. In Germany it is possible to control or
trace the history of infected people by local health institutions 
if the seven-days incidence has a value less
than 50. But at the end of December 2020 and the begin of January 2021 the averaged
incidence is about 140, and in some hotspot federal states like Saxony greater than 300.
If the social and economical life should be sustained there are
several possibilities to transmit the COVID-19 virus anyhow. The following ones
should be mentioned:
\begin{itemize}
	\item Commuters and employers on the way to there office or to there position of employment especially including medical and nursing staff.
	\item Pupils and teachers in schools and on the way to school and in the school.
	\item People buying every day necessities using shopping centers. 
	\item Postmen, suppliers and deliverers.
\end{itemize}
All these activities take place during the so called lockdown in Germany with
the result of ongoing propagation of the pandemic. Also the unavailable center
of power in the decentralized federal state Germany. This leads often to solo
efforts of some federal states.

\begin{figure}[h]
	\begin{minipage}[t]{0.49\textwidth}
		\includegraphics[width=7.2cm]{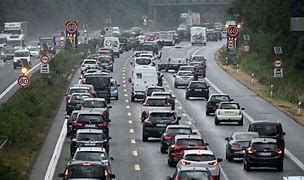}
		\caption{\label{fig0a} German lockdown}
	\end{minipage}
	\hfill
	\begin{minipage}[t]{0.49\textwidth}
		\includegraphics[width=7.3cm]{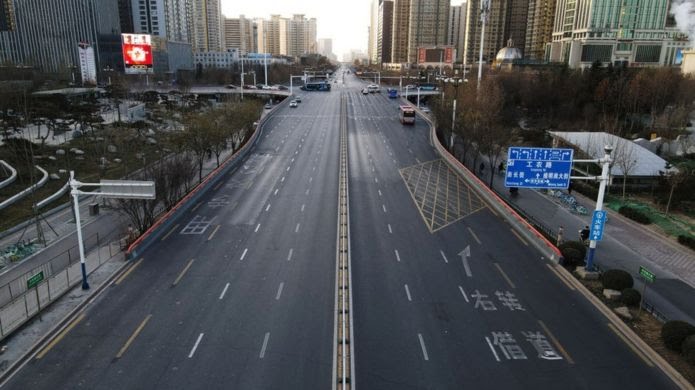}
		\caption{\label{fig0b} Chinese lockdown}
	\end{minipage}
\end{figure}

From authoritarian countries like China or Singapore with a quite different
civilization and other cultural traditions than the German ones for example it is known,
that the virus propagation could be stopped with very rigorous measures
like the strict prohibition of the social and economic life (see \ref{fig0b}\footnote{From gettyimages}), i.e. activities
which mentioned above are absolutely forbidden.

This is inconceivable in countries like Germany, Austria, the Netherlands
or other so called democratic states with a western understanding of
freedom and self-determination (see \ref{fig0a}).
But in the consequences of such a western life style they have to live with 
a more or less consecutive activity of the COVID-19 pandemic. And that
is the reason for the following trial to describe one aspect of the
pandemic by a diffusion model. In another context a similar model was discussed
in \cite{braa}.

In the following diffusion concept the seven-days incidence should be denoted
by $s$ and $s$ should serve as the quantity which will be influenced by it's
gradients between different levels of incidence in the federal states of Germany.
At the the mathematical model of diffusion leads to a partial differential equation
for the considered quantity (here $s$)
\begin{equation}\label{eq1}
\frac{\partial s}{\partial t} = \nabla\cdot (D \nabla s) + q \quad \mbox{in}\; [t_0,T]\times\Omega, 
\end{equation}
where $\Omega\subset \R^2$ is the region which will be investigated, for example
the national territory of Germany, $D$ is a diffusion coefficient, depending on the
locality $x\in \Omega$, $[t_0,T]$ is the time interval of interest, and $q$ is a term
which describes sources or sinks of possible infections.

Beside the equation \eqref{eq1} one needs initial conditions for $s$, for example
\begin{equation}\label{eq2}
   s(x,t_0) = s_0 (x),\quad x\in \Omega
\end{equation}
and boundary conditions
\begin{equation}\label{eq3}
   \alpha s +  \beta \nabla s\cdot\vec{n} = \gamma  \quad \mbox{in}\; [t_0,T]\times\partial\Omega \,,
\end{equation}
where $\alpha,\beta$ and $\gamma$ are real coefficients and by $\partial\Omega =: \Gamma$
the boundary of the region $\Omega$ will be denoted. $\nabla_n s = \nabla s\cdot \vec{n}$
is the directional derivative of $s$ in the direction of the outer normal vector $\vec{n}$ on $\Gamma$.
The choice of $\alpha = 0$, $\beta = 1$ and $\gamma = 0$ for example
leads to the homogeneous Neumann boundary condition
\begin{equation}\label{eq4}
    \nabla_n s = 0\;,
    \end{equation}
which means no import of $s$ at the boundary $\Gamma$. In other words, \eqref{eq4} describes
closed borders to surrounding countries outside $\Omega$.

The diffusion coefficient function $D: \Omega \to \R$ is responsible for the 
intensity or velocity of the diffusion process. From fluid or gas dynamics one knows from \cite{cus} the
formula 
\[    D = \frac{2}{3} \bar{v} l   \]
with the averaged particle velocity $\bar{v}$ and the mean free path $l$. 
The application of this ansatz to the movement of people in certain areas requires
assumptions for $\bar{v}$ and $l$. If we consider a circular or quadratic region
with the area $A$ and a number of inhabitants $N$ who are distributed equally $l$ 
could be approached by
\[   l = \frac{\sqrt{A}}{\sqrt{N}} \;. \]
For the velocity $\bar{v}$ we assume $\bar{v} = 5000 \frac{km}{day}$ (\cite{bont}).
Because of the different areas and numbers of inhabitants of the federal states of Germany
$D$ will be a local depending non-constant function.

If there are no sources or sinks for $s$, i.e. $q = 0$, and the borders are closed
which means the boundary condition \eqref{eq4}, the initial boundary value problem
\eqref{eq1}, \eqref{eq2}, \eqref{eq4} has the steady state solution
\begin{equation}\label{eq5}
s_{st} = \frac{\int_\Omega s_0 (x)\, dx}{\int_\Omega dx} = \mbox{const.}\;.
\end{equation}
This is easy to verify and
this property is characteristic for diffusion processes which tend to an equilibrium.
It is quite complicated to model the source-sink function $q$ in an appropriate kind.
$q$ depends of the behavior of the population and the health policy of the different
federal states. It's only possible to work with very coarse guesses. It is known that
the people in Schleswig-Holstein is exemplary with respect to the recommendations to
avoid infection with the COVID-19 virus and this means $q < 0$. On the other hand it is
known from Saxony the many people belief there is not a jot of truth in the pandemic, which
means $q>0$ for a long time (now the government of Saxony changed the policy which leads
to $q < 0$).

But regardless of these uncertainties one can get information about the pandemic
propagation for example the influence of hotspots of high incidences (Saxony) to regions with
low incidences (South of Brandenburg) for example.

\section{Data of the different federal states of Germany}
At the beginning of the year 2021 (14th of January) the Robert-Koch-Institut which
is responsible for the daily COVID-19 data collection published the seven-days incidence
data (of January the 14th, 2021, see \cite{rki}) summarized in table \ref{t1}.

The values of table \ref{t1} are used as initial data for the function
$s_0$ of \eqref{eq2}.

As a base for the determination of the diffusion coefficient function we use the
data of table \ref{t1}.

\begin{table}[htb]
	\begin{center}
		\begin{tabular}{|l|r|r|r|r|}\hline
			       states & 7-days incidence & density & inhibitants & area  \\ \hline\hline
			Schleswig-Holstein & 92 & 183 & 2904 & 15804\\
			Hamburg & 115 & 2438 & 1847 & 755\\
			Mecklenburg-West Pomerania & 117 & 69 & 1608 & 23295\\
			Lower Saxony & 100 & 167 & 7994 & 47710\\
			Brandenburg & 212 & 85 & 2522 & 29654\\
			Berlin & 180 & 4090 & 3669 & 891\\
			Bremen & 84 & 1629 & 681 & 419\\
			Saxony-Anhalt & 241 & 109 & 2195 & 20454\\
			Thuringia & 310 &132 & 2133 & 16202\\
			Saxony & 292 & 221 & 4072 & 18450\\
			Bavaria & 160 &185 & 13125 & 70542\\
			Baden-Wuerttemberg & 133 & 310 & 11100 & 35784\\
			North Rhine-Westphalia & 131 & 526 & 17947 & 34112\\
			Hesse & 141 & 297 & 6288 & 21116\\
			Saarland & 160 & 385 & 987 & 2571\\
			Rhineland-Palatinate & 122 & 206 & 4094 & 19858\\
			Munic & 156 & 4700 & 1540 & 310\\ \hline
		\end{tabular}
		\caption{\label{t1} 7-days incidence, people density $[/km^2]$, inhibitants $[/100000]$, area of the federal states of Germany $[km^2]$}
	\end{center}
\end{table}

The unit of the diffusion function $D$ will be $[km^2/day]$. Eventual sources or
sinks will be gauged in $[/day]$. The incidence $s$ is dimensionless.

\section{The numerical solution of the initial boundary value problem \eqref{eq1},\eqref{eq2},\eqref{eq4}}
Based on the subdivision of $\Omega$ (area of Germany)
into finite rectangular cells $\omega_j,\,j\in I_\Omega$ and $\Omega = \cup_{j \in I_\Omega} \omega_j$ the equation \eqref{eq1} will be spatial discretized
with a finite volume method ($I_\Omega$ is the index set of the finite volume cells)
. Together with the discrete boundary condition
\eqref{eq4} we get a semi-discrete system continuous in time
\begin{equation}\label{eq6}
\frac{\partial s_j}{\partial t} = \nabla_h \cdot (D \nabla_h s_j ) + q_j\;,\; j \in I_\Omega,
\end{equation}
where the index $h$ means the discrete versions of the $\nabla$-operator. 

The time discretization is done with an implicit Euler scheme. This allows us to
work without strict restrictions for the choice of the discrete time-step $\Delta_t$.
At every time-level it is to solve the linear equation system
\begin{equation}\label{eq7}
\frac{1}{\Delta_t} s_j^{n+1} - \nabla_h \cdot (D \nabla_h s_j^{n+1}) = \frac{1}{\Delta_t} s_j^n + q_j  \;,\; j\in I_\Omega\;,
\end{equation}
for $n = 0,\dots N,\, N = (T- t_0)/\Delta_t$. $s_j^0$ was set to the incidence
$s_0 (x)$ for $x\in \omega_j$, $j=1,\dots,I_\Omega$.
The solution of equation system \eqref{eq7} for a certain time-level $n$
is done with an iterative method.

\section{Numerical simulation results}
\begin{figure}[h]
	\begin{minipage}[t]{0.49\textwidth}
		\includegraphics[width=8.7cm]{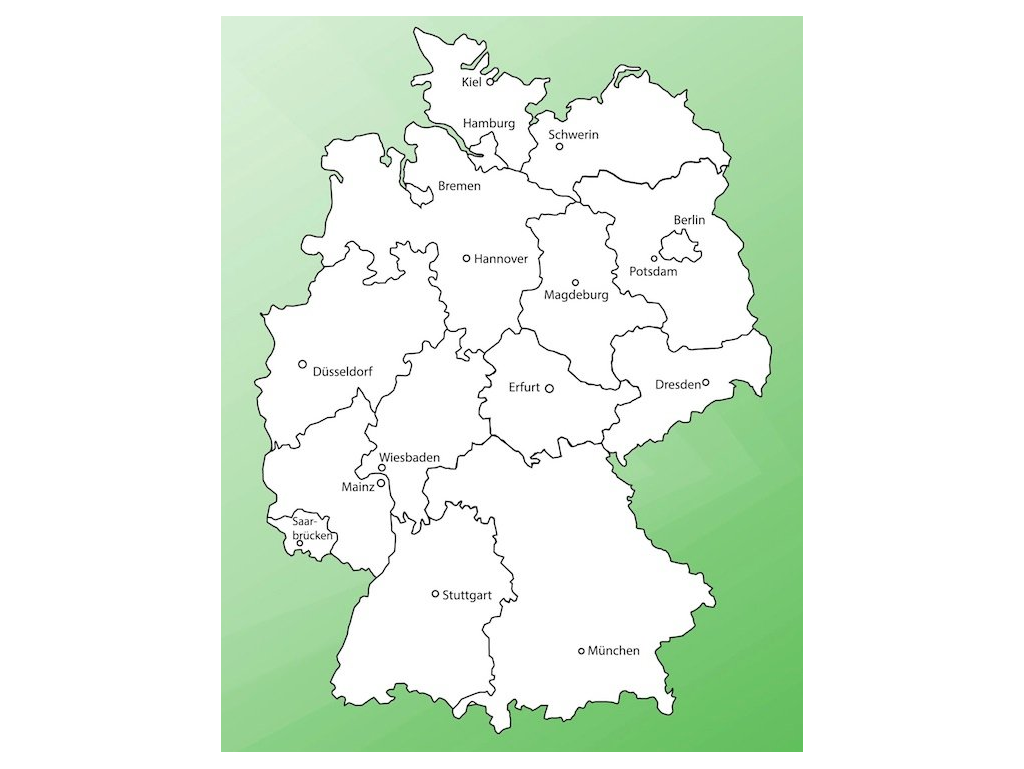}
		\caption{\label{fig01} Germany map}
	\end{minipage}
	\hfill
	\begin{minipage}[t]{0.49\textwidth}
        \includegraphics[width=6.7cm,height=8cm]{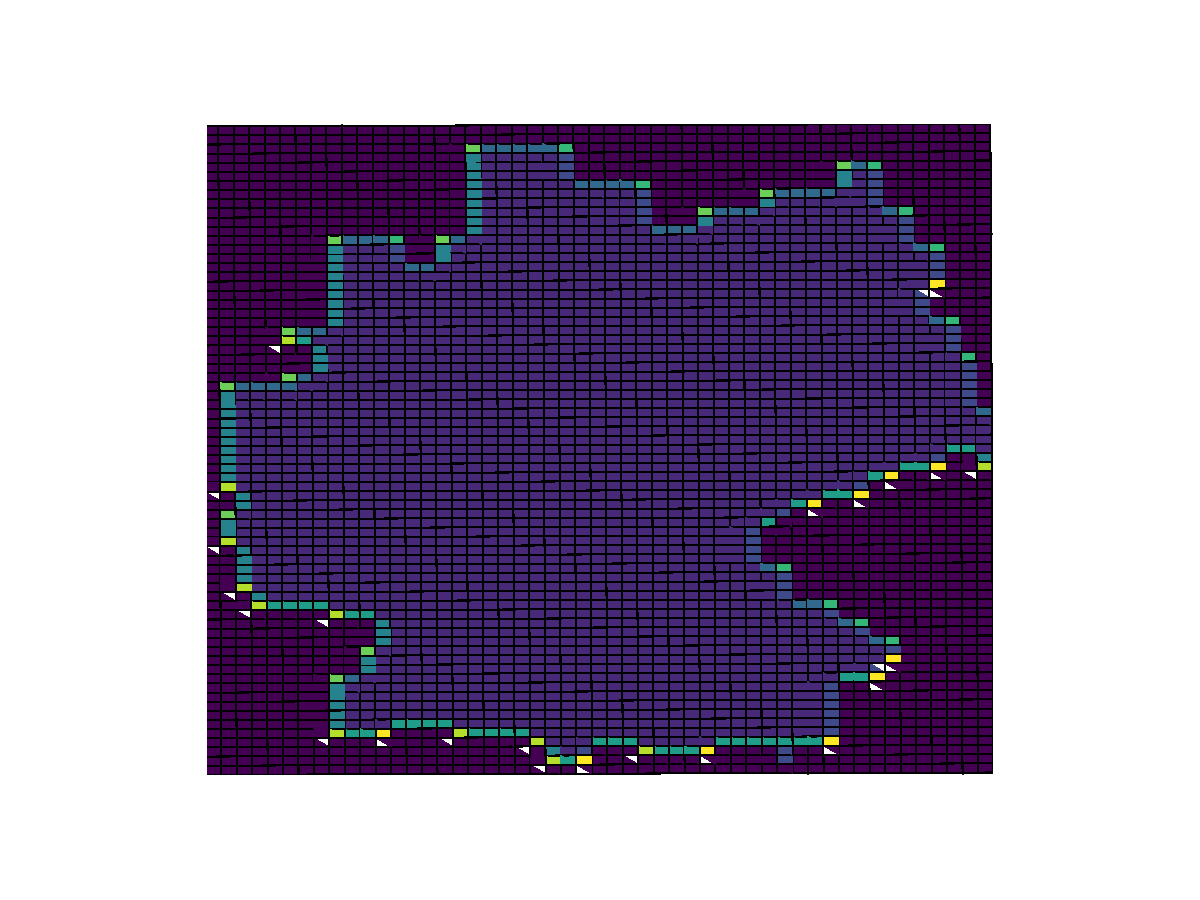}
		\caption{\label{fig02} Coarse contour of $\Omega$ and it's discretization}
	\end{minipage}
\end{figure}

In fig. \ref{fig02} the region $\Omega$ is adumbrated. The size of
finite volume cells is $\Delta_x \times \Delta_y = (8km\times 8km)$.
First we start with the case $q = 0$. $\Delta_t$ was set to one day.
In the following figs. \ref{fig11}-\ref{fig32} the initial state and the results of the diffusion
process for the development of the seven-days incidence 
after 50 and 120 time-steps (days) are shown. The left figs. show a view from
west to east, and the right figs. show the view from north to south.
The initial state is a piece-wise
constant function with values of the seven-days incidence of the 16 federal states
where we consider munic as a town with over a million inhabitants separately (it was
excluded from Bavaria).
\begin{figure}[h]
\begin{minipage}[t]{0.49\textwidth}
		\includegraphics[width=8.7cm]{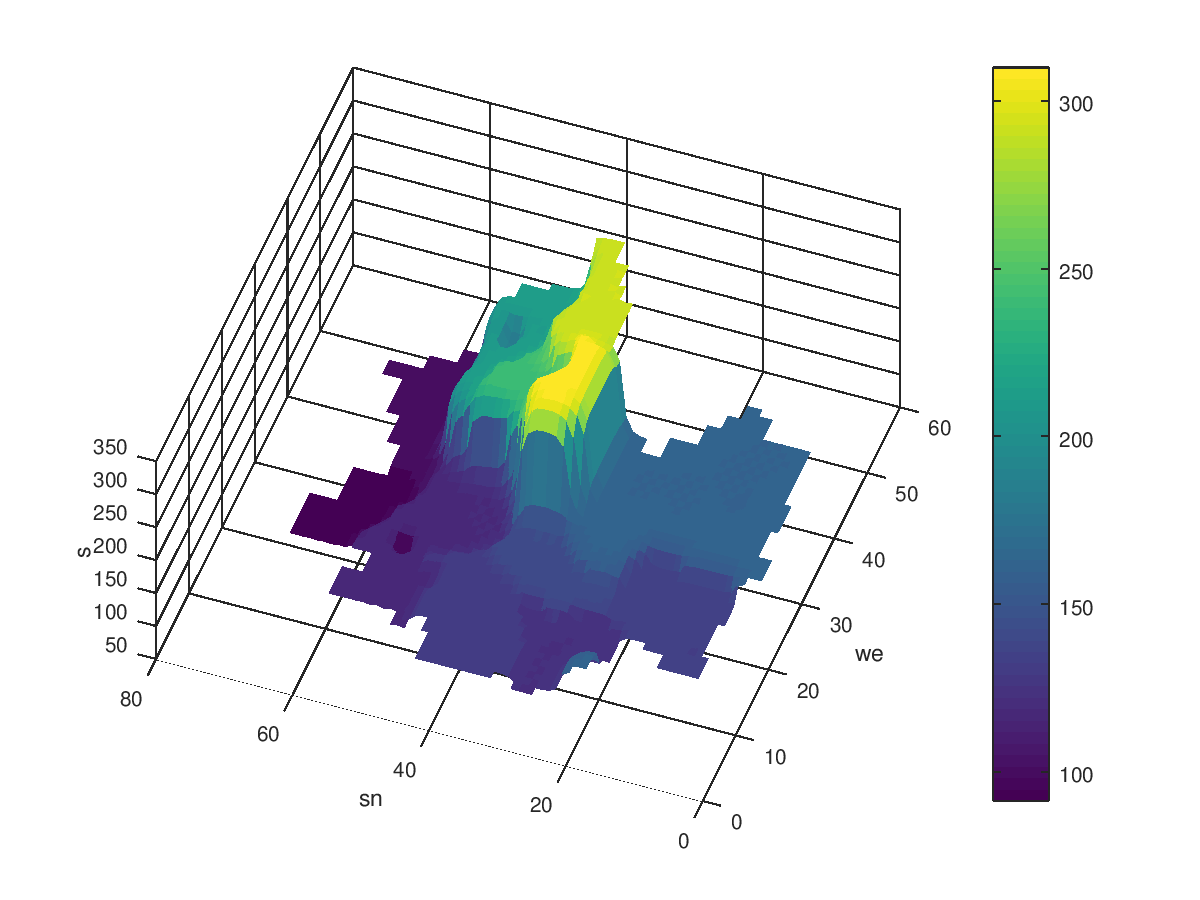}
\caption{\label{fig11} Initial distribution of $s$, $n=0$, $q=0$}
\end{minipage}
\hfill
\begin{minipage}[t]{0.49\textwidth}
		\includegraphics[width=8.7cm]{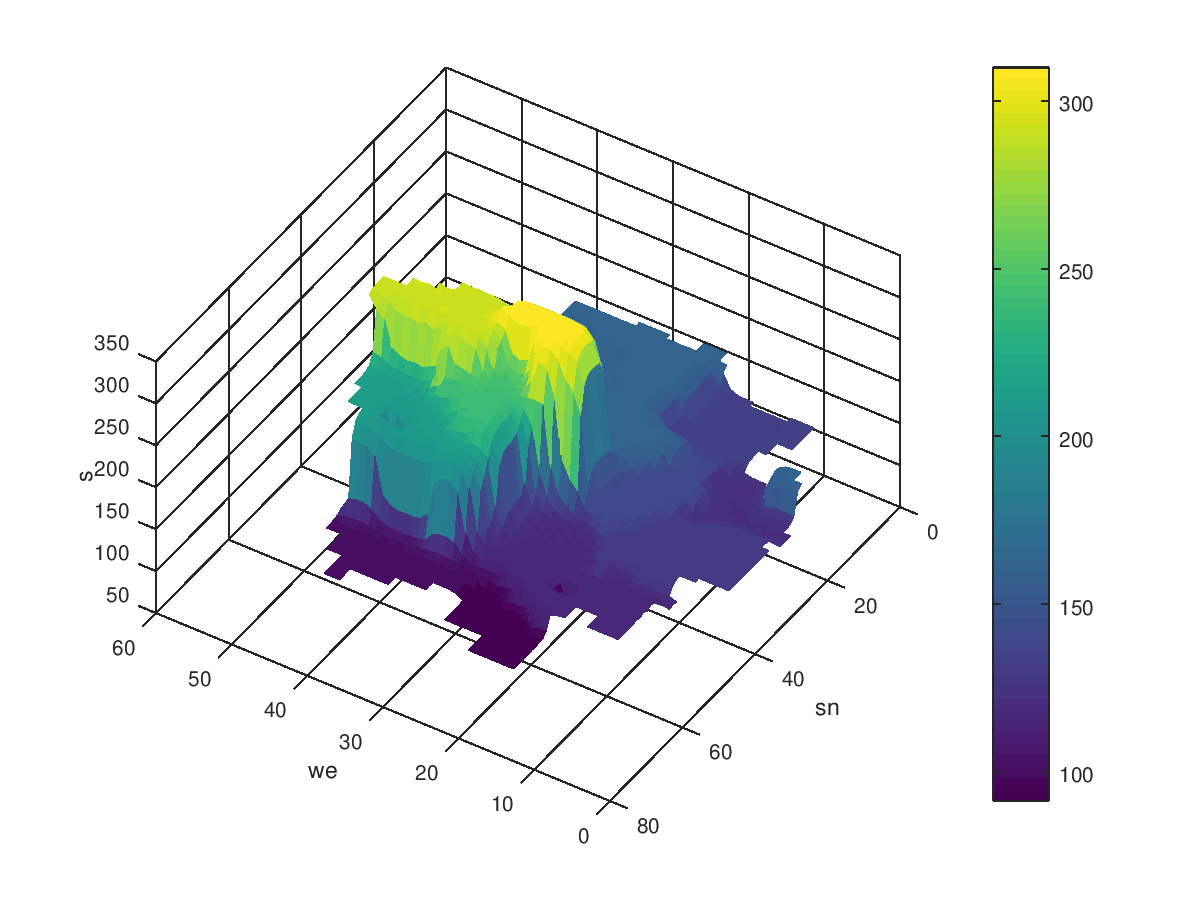}
\caption{\label{fig12} Initial distribution of $s$, $n=0$, $q=0$}
\end{minipage}
\end{figure}
\begin{figure}[h]
	\begin{minipage}[t]{0.49\textwidth}
		\includegraphics[width=8.7cm]{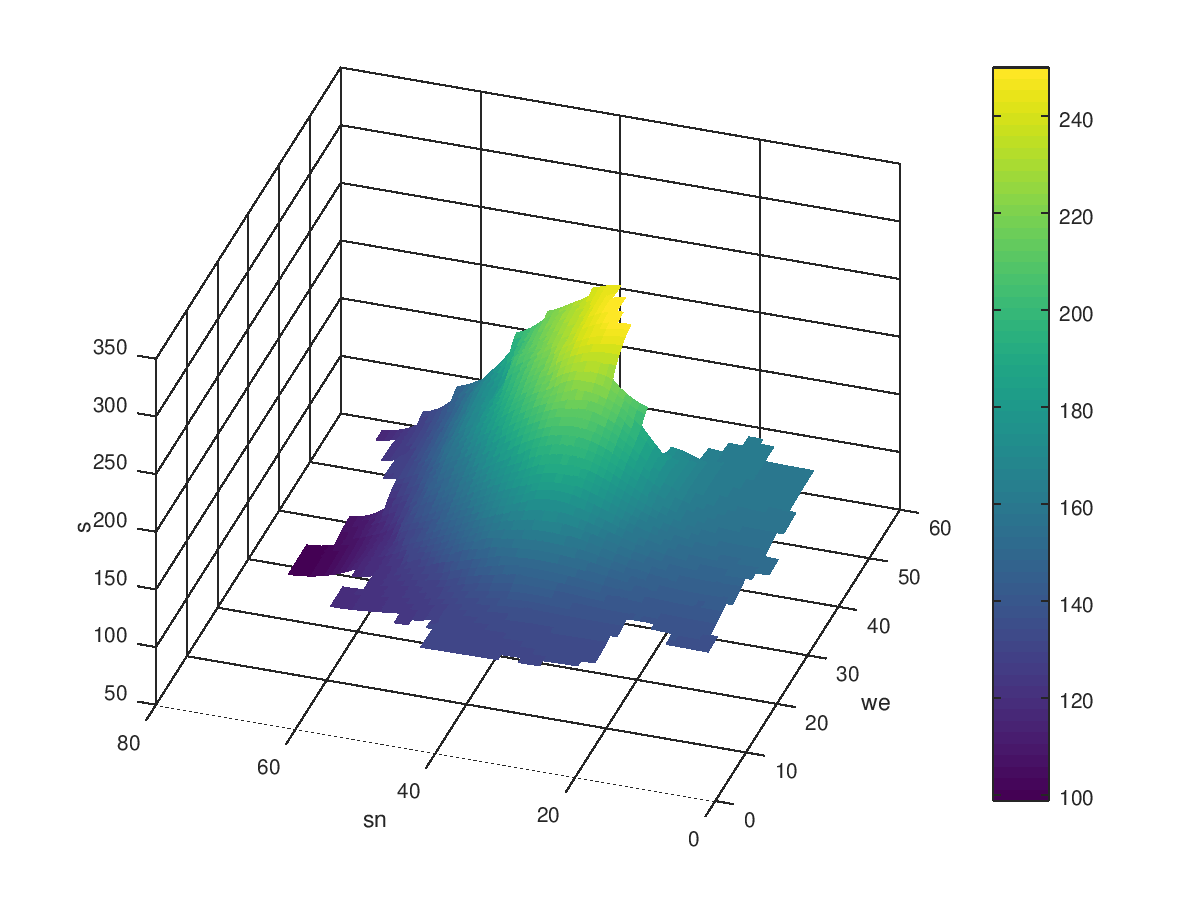}
		\caption{\label{fig21} $s$ after 100 time-steps, $q=0$}
	\end{minipage}
	\hfill
	\begin{minipage}[t]{0.49\textwidth}
		\includegraphics[width=8.7cm]{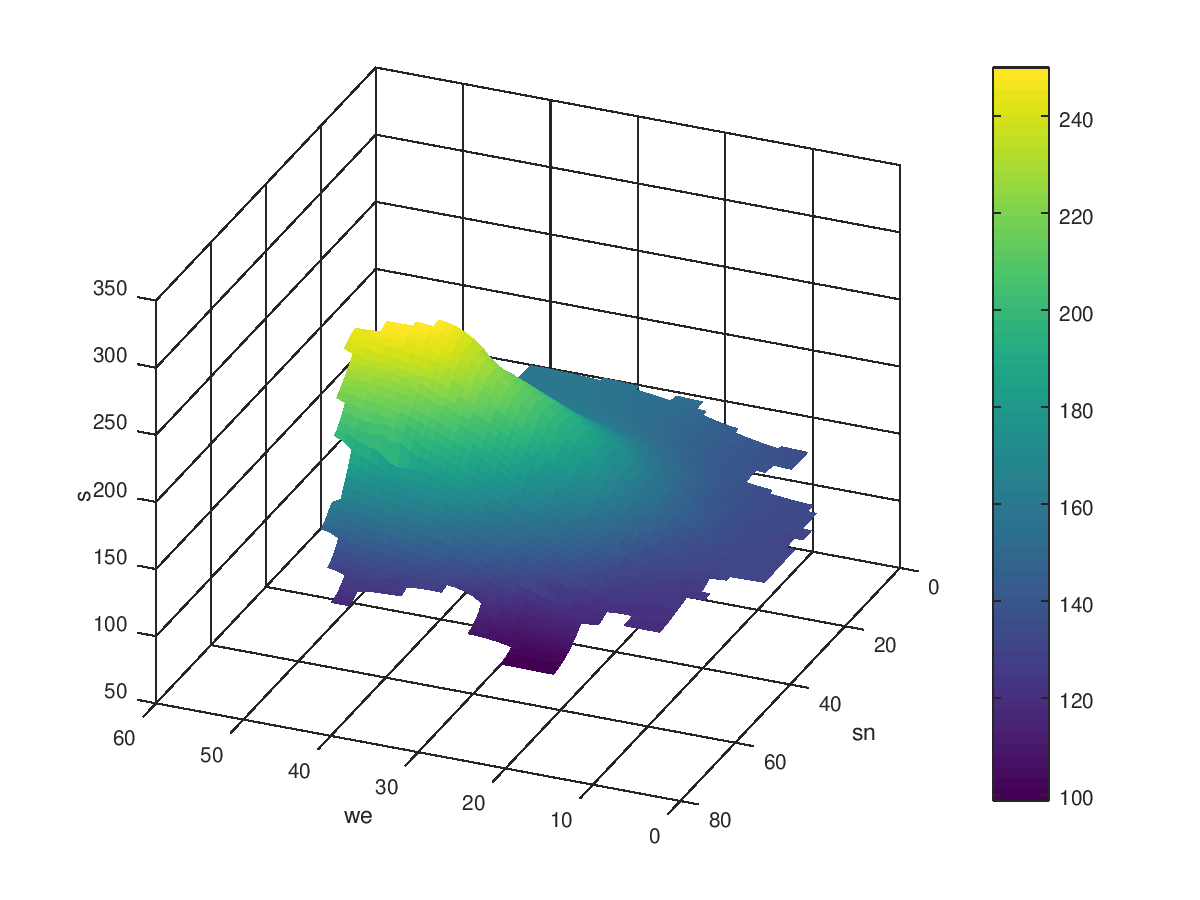}
		\caption{\label{fig22} $s$ after 100 time-steps, $q=0$}
	\end{minipage}
\end{figure}
\begin{figure}[h]
	\begin{minipage}[t]{0.49\textwidth}
		\includegraphics[width=8.7cm]{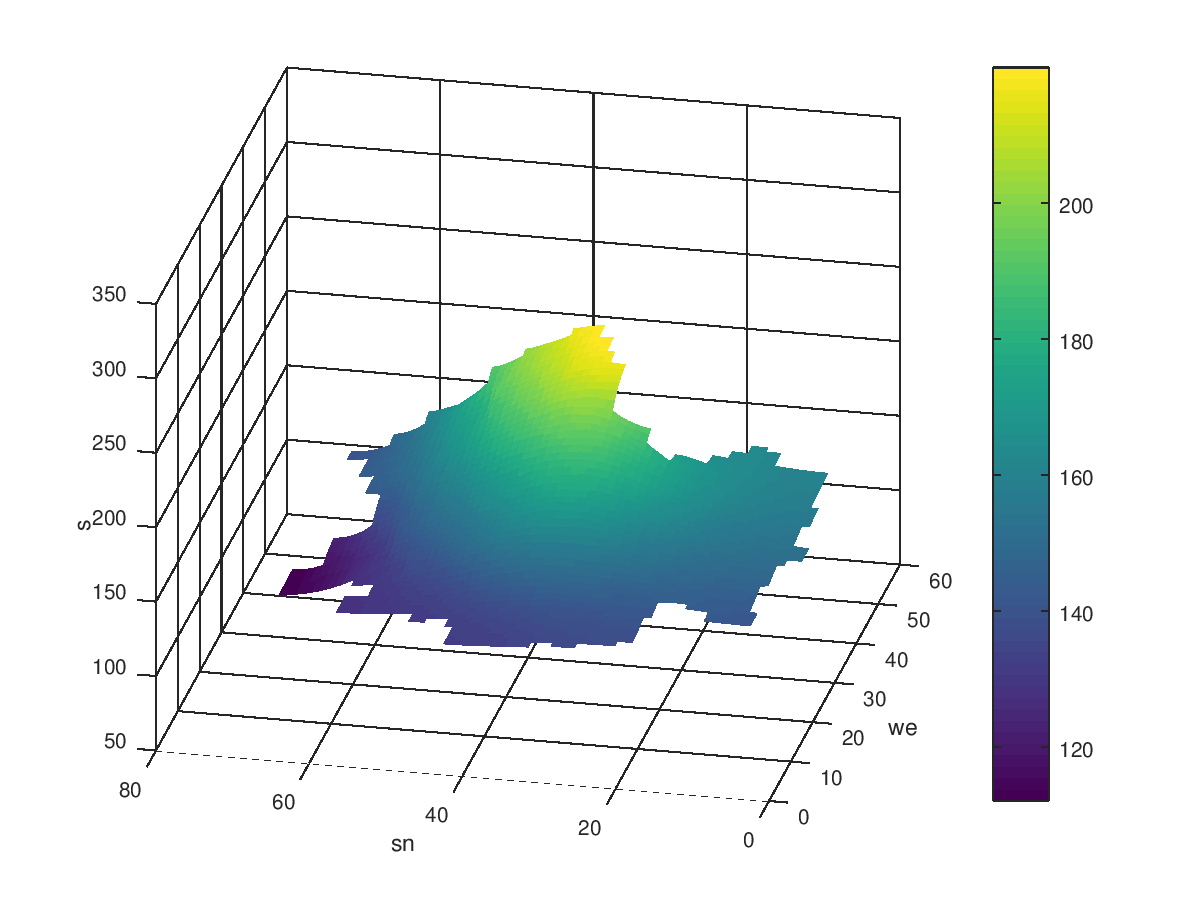}
		\caption{\label{fig31} $s$ after 200 time-steps, $q=0$}
	\end{minipage}
	\hfill
	\begin{minipage}[t]{0.49\textwidth}
		\includegraphics[width=8.7cm]{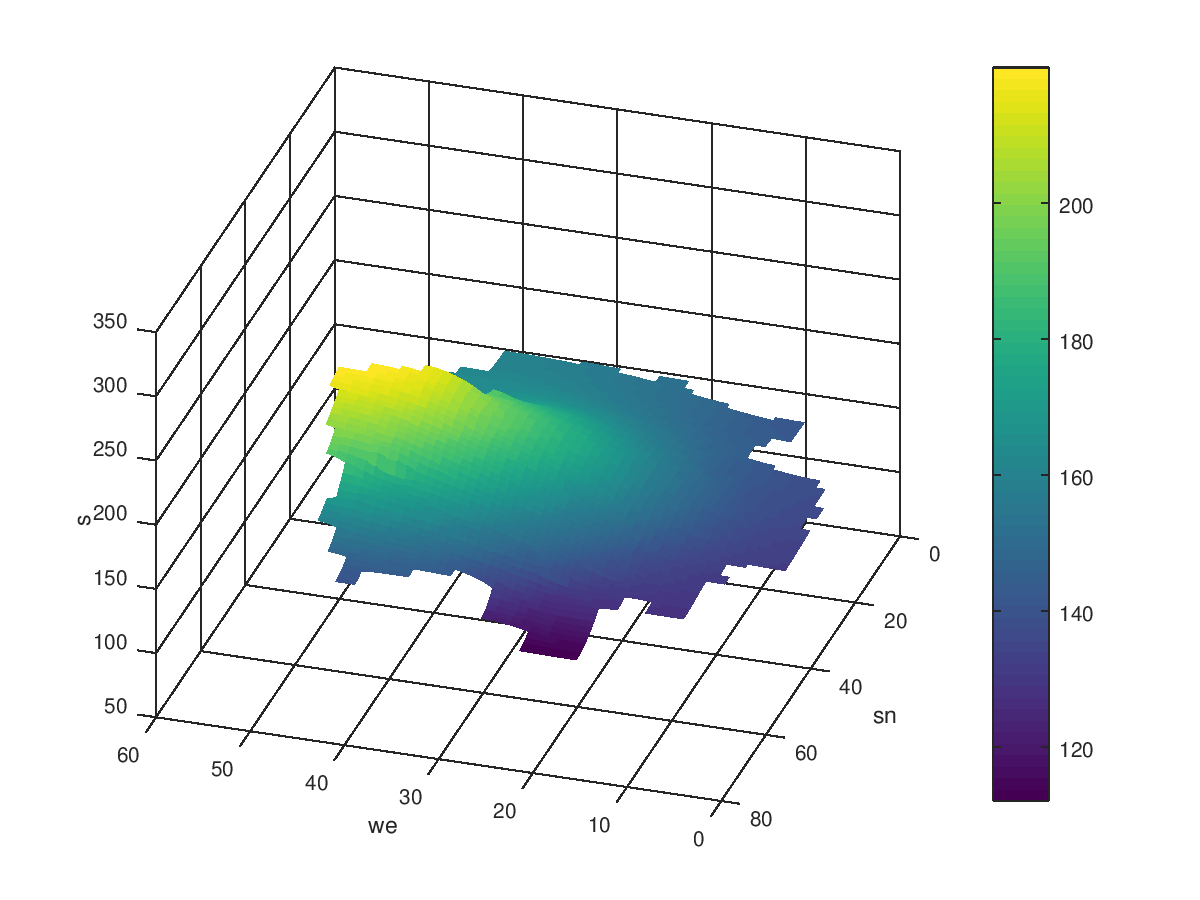}
		\caption{\label{fig32} $s$ after 200 time-steps, $q=0$}
	\end{minipage}
\end{figure}
Especially in the border regions (Saxony - Brandenburg, Saxony - Bavaria,
Saxony - Thuringia) one can observe a transfer of incidence from the high
level incidence of Saxony to the neighbored federal states. Also the high incidence
level of Berlin was transferred to the surrounding area of Brandenburg.
The north states with a low incidence level were only influenced by
the other states weakly.

With the parameters $\alpha$, $\beta$ and $\gamma$ of the boundary condition \eqref{eq3} it's possible to describe several situations
at the borders of the boundary
$\Gamma$ of $\Omega$. $\alpha = 0$, $\beta = -D$ and $\gamma \neq 0$ describes
a flux through the border. In the next example such a scenario will be
used to describe the way home of infected people from Austria to Bavaria.
The boundary condition at the border crossing reads as
\[  -D \nabla s\cdot\vec{n} = \gamma \;.\]
The initial state $s_0$ was the same as in the example above. $\gamma > 0$ 
means an ''inflow'' of infected people, $\gamma < 0$ a loss of infected people
($\gamma = 0$ describes a closed border).
In the following figures \ref{fig41}, \ref{fig42} the case $\gamma = 0$ is compared to the case $\gamma = 50$ ($km/day$). At the south boundary of Bavaria one can
observe the increase of $s$ caused by the flux of $s$ from Austria to Bavaria.
\begin{figure}[h]
	\begin{minipage}[t]{0.49\textwidth}
		\includegraphics[width=8.7cm]{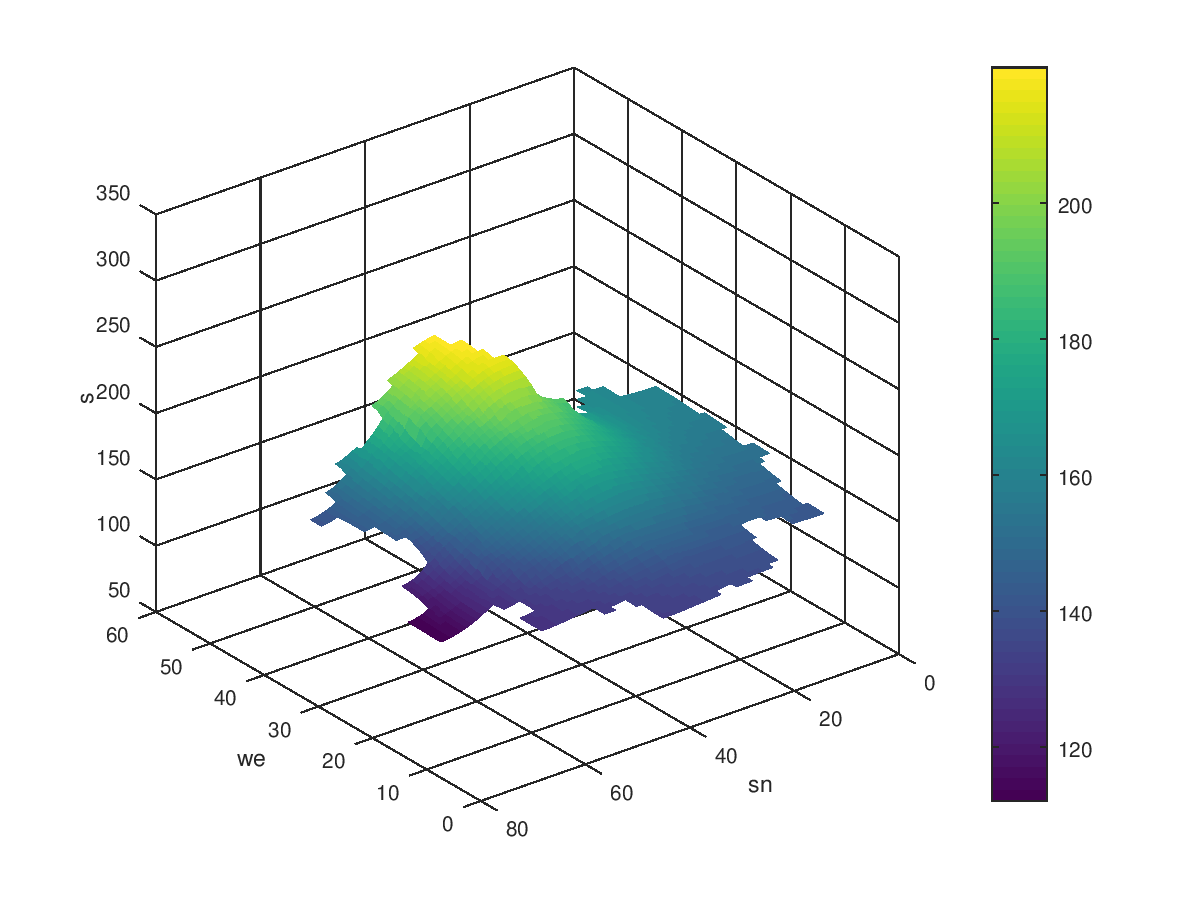}
		\caption{\label{fig41} $s$ after 200 time-steps, $\gamma= 0 km/day$, $q=0$}
	\end{minipage}
	\hfill
	\begin{minipage}[t]{0.49\textwidth}
		\includegraphics[width=8.7cm]{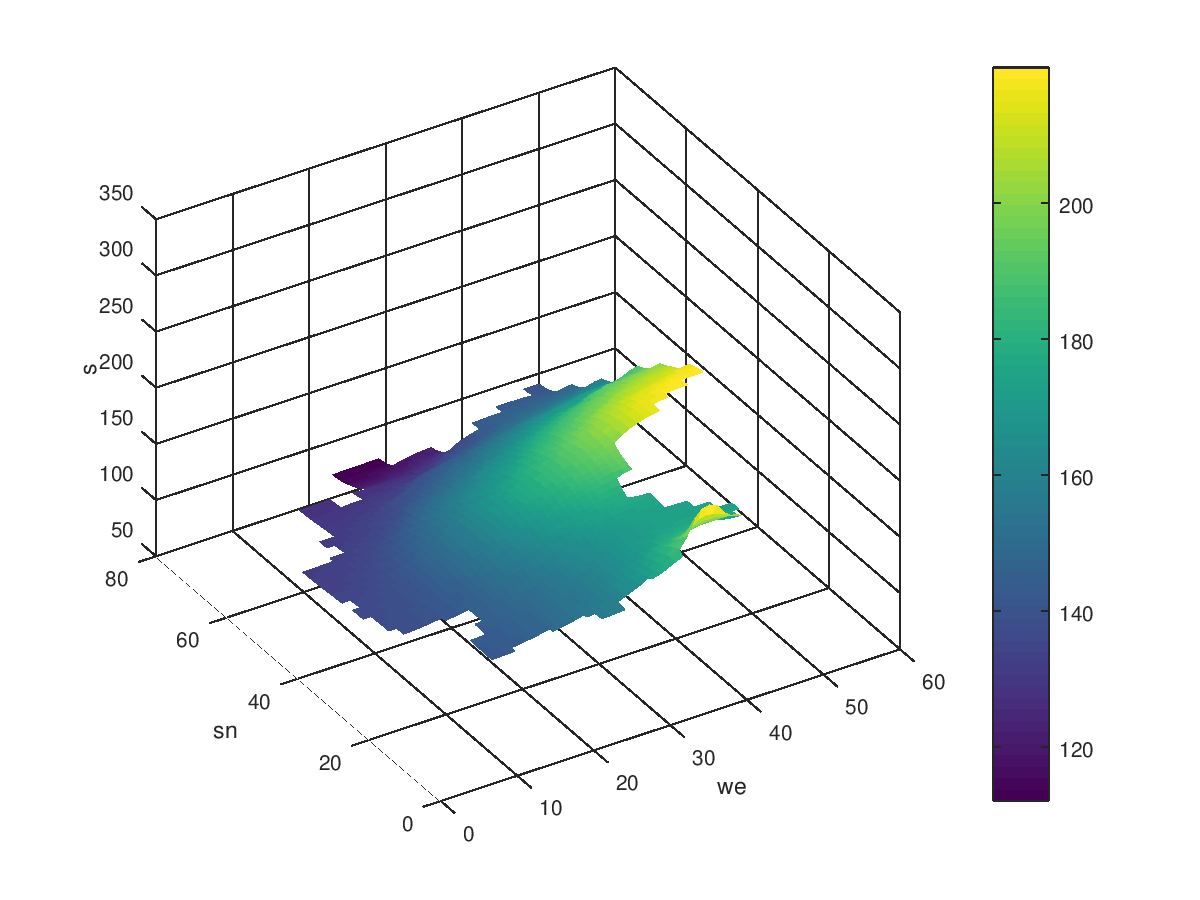}
		\caption{\label{fig42} $s$ after 200 time-steps, $\gamma = 50 km/day$, $q=0$}
	\end{minipage}
\end{figure}
\begin{figure}[htb]
        \begin{minipage}[t]{0.49\textwidth}
                \includegraphics[width=8.7cm]{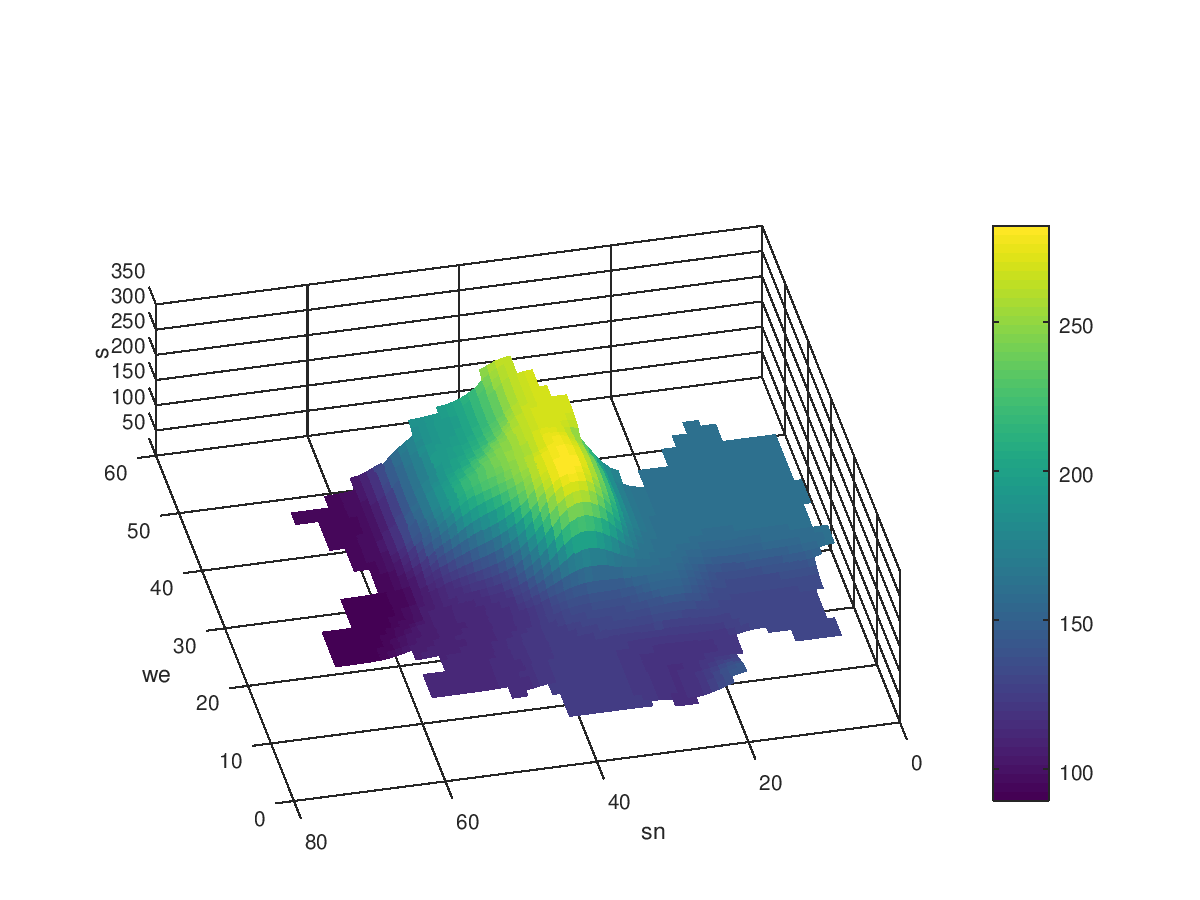}
                \caption{\label{fig61} $s$ after 10 time-steps, $q\neq 0$, based on table \ref{t2}}
        \end{minipage}
        \hfill
        \begin{minipage}[t]{0.49\textwidth}
                \includegraphics[width=8.7cm]{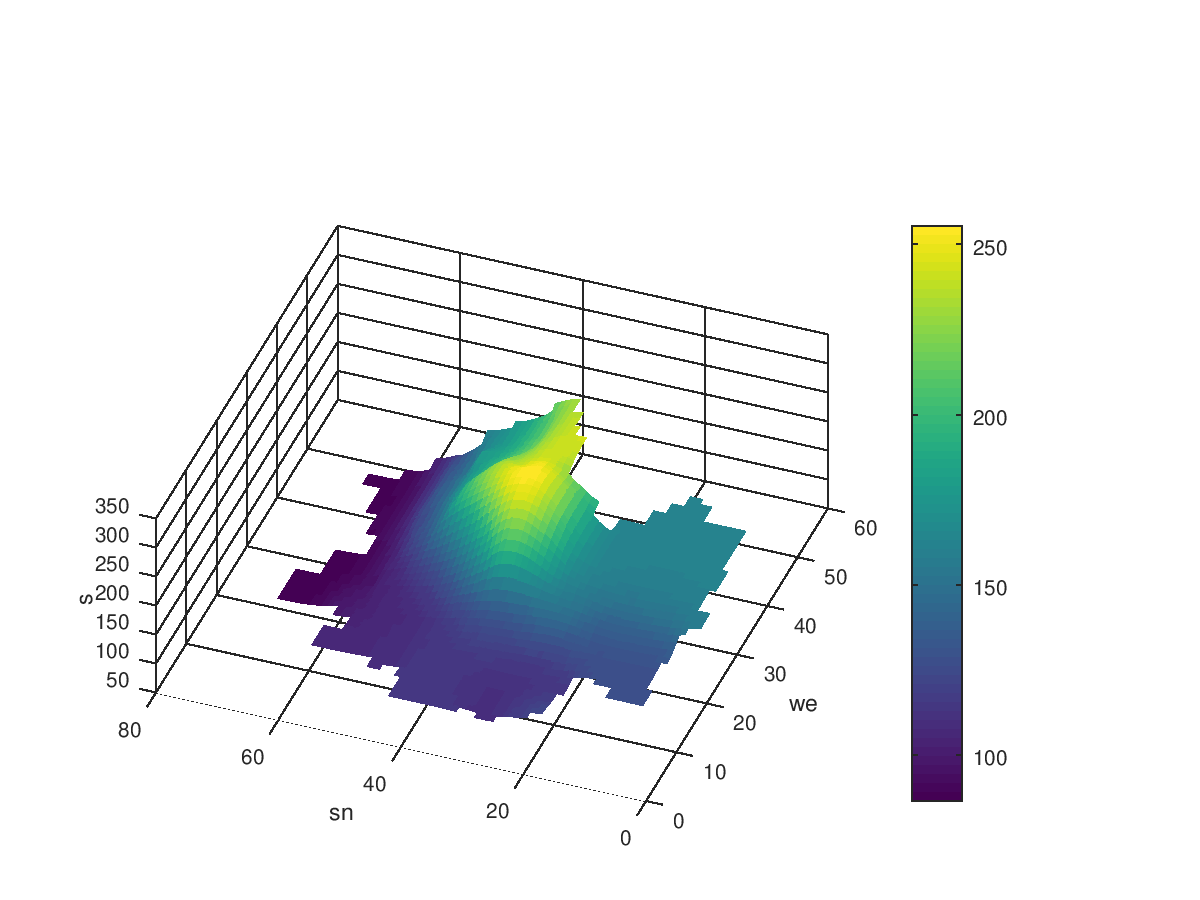}
                \caption{\label{fig62} $s$ after 20 time-steps, $q\neq 0$, based on table \ref{t2}}
        \end{minipage}
\end{figure}
\begin{figure}[htb]
        \begin{minipage}[t]{0.49\textwidth}
                \includegraphics[width=8.7cm]{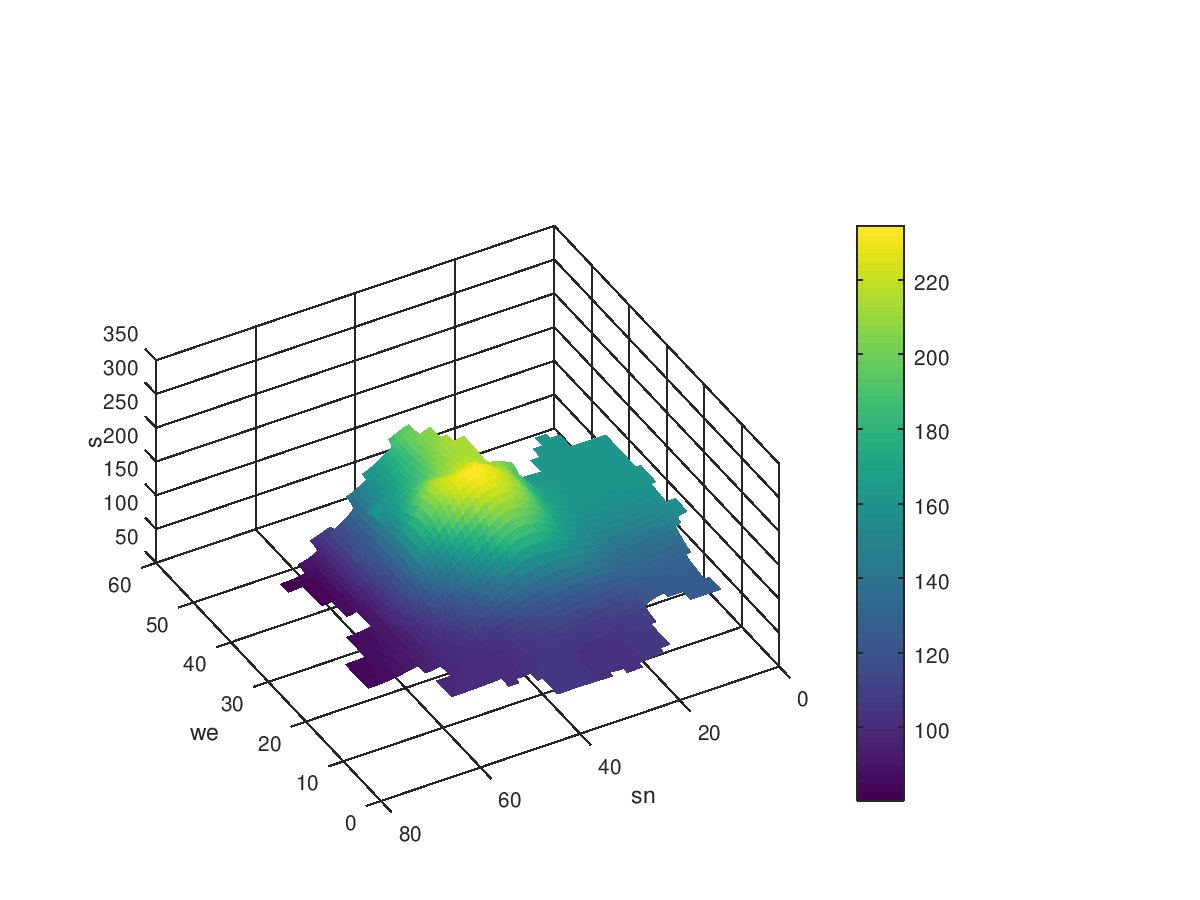}
                \caption{\label{fig63} $s$ after 30 time-steps, $q\neq 0$, based on table \ref{t2}}
        \end{minipage}
        \hfill
        \begin{minipage}[t]{0.49\textwidth}
                \includegraphics[width=8.7cm]{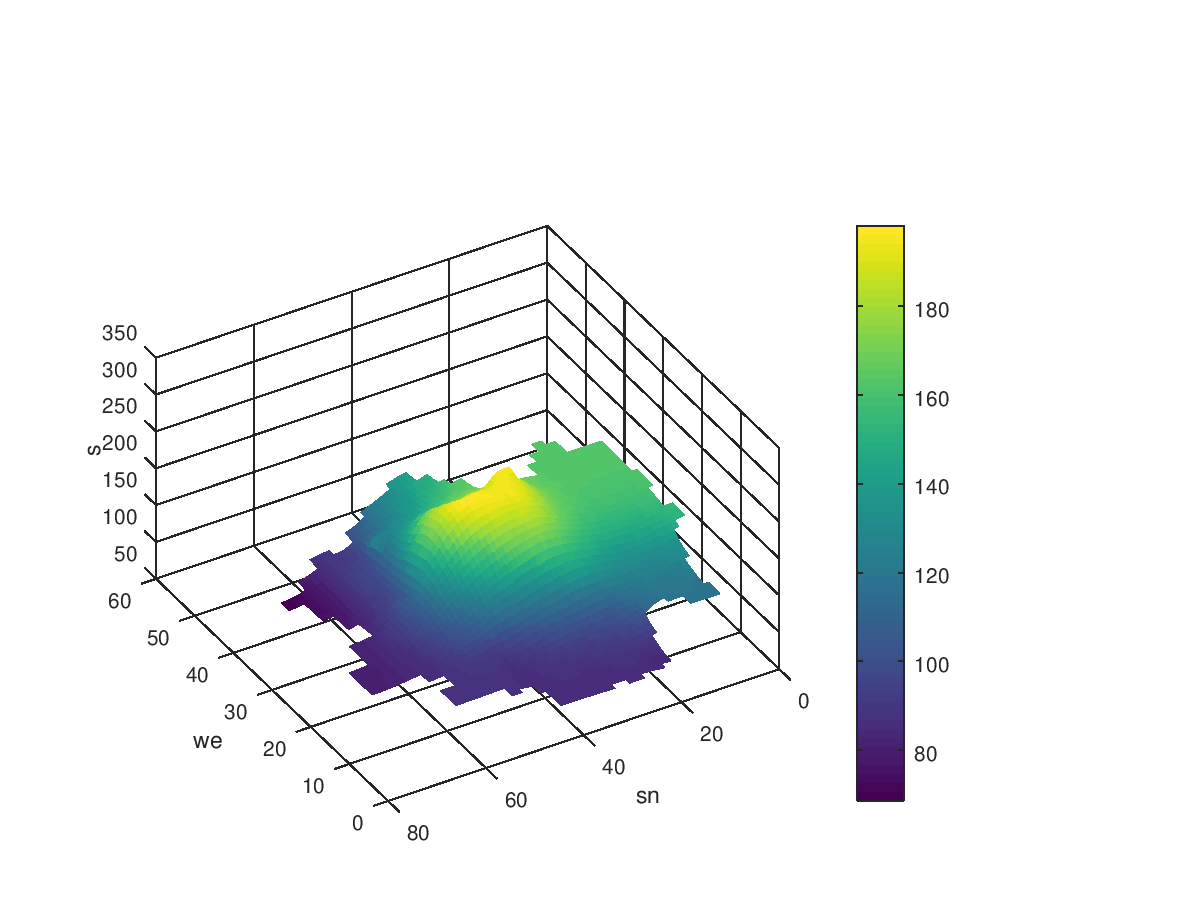}
                \caption{\label{fig64} $s$ after 50 time-steps, $q\neq 0$, based on table \ref{t2}}
        \end{minipage}
\end{figure}

\begin{table}[htb]
        \begin{center}
                \begin{tabular}{|l|r|r|r|r|}\hline
                               states & 12.1.2021 & 13.1.2021 & 14.1.2021 & $\Delta s$ per day\\ \hline\hline
                        Schleswig-Holstein & 98 & 94 & 92 & -3  \\
                        Hamburg & 127 & 120 & 115 & -1  \\
                        Mecklenburg-West Pomerania & 129 & 122 & 117 & -6 \\
                        Lower Saxony & 114 & 108 & 100 & -7 \\
                        Brandenburg & 258 & 230 & 212 & -23 \\
                        Berlin & 178 & 184 & 180 & 1 \\
                        Bremen & 86 & 84 & 84 & -1 \\
                        Saxony-Anhalt & 238 & 232 & 241 & 1,5 \\
                        Thuringia & 326 & 324 & 310 & -8 \\
                        Saxony & 342 & 304 & 292 & -25 \\
                        Bavaria & 159 & 148 & 169 & 0,5  \\
                        Baden-Wuerttemberg & 139 & 130 & 133 & -3 \\
                        North Rhine-Westphalia & 149 & 142 & 131 & -9 \\
                        Hesse & 157 & 150 & 141 & -8  \\
                        Saarland & 184 & 176 & 160 & -12 \\
                        Rhineland-Palatinate & 139 & 132 & 132 & -8,5 \\
                        Munic & 157 & 156 & 156 & 0 \\ \hline
                \end{tabular}
                \caption{\label{t2} 7-days incidence, guessed changed $s$ per day $[/day]$}
        \end{center}
\end{table}
\begin{figure}[thb]
        \begin{minipage}[t]{0.49\textwidth}
                \includegraphics[width=8.2cm]{inci_sn_200.png}
                \caption{\label{fig51} $s$ after 200 time-steps, $q = 0$}
        \end{minipage}
        \hfill
        \begin{minipage}[t]{0.49\textwidth}
                \includegraphics[width=8.2cm]{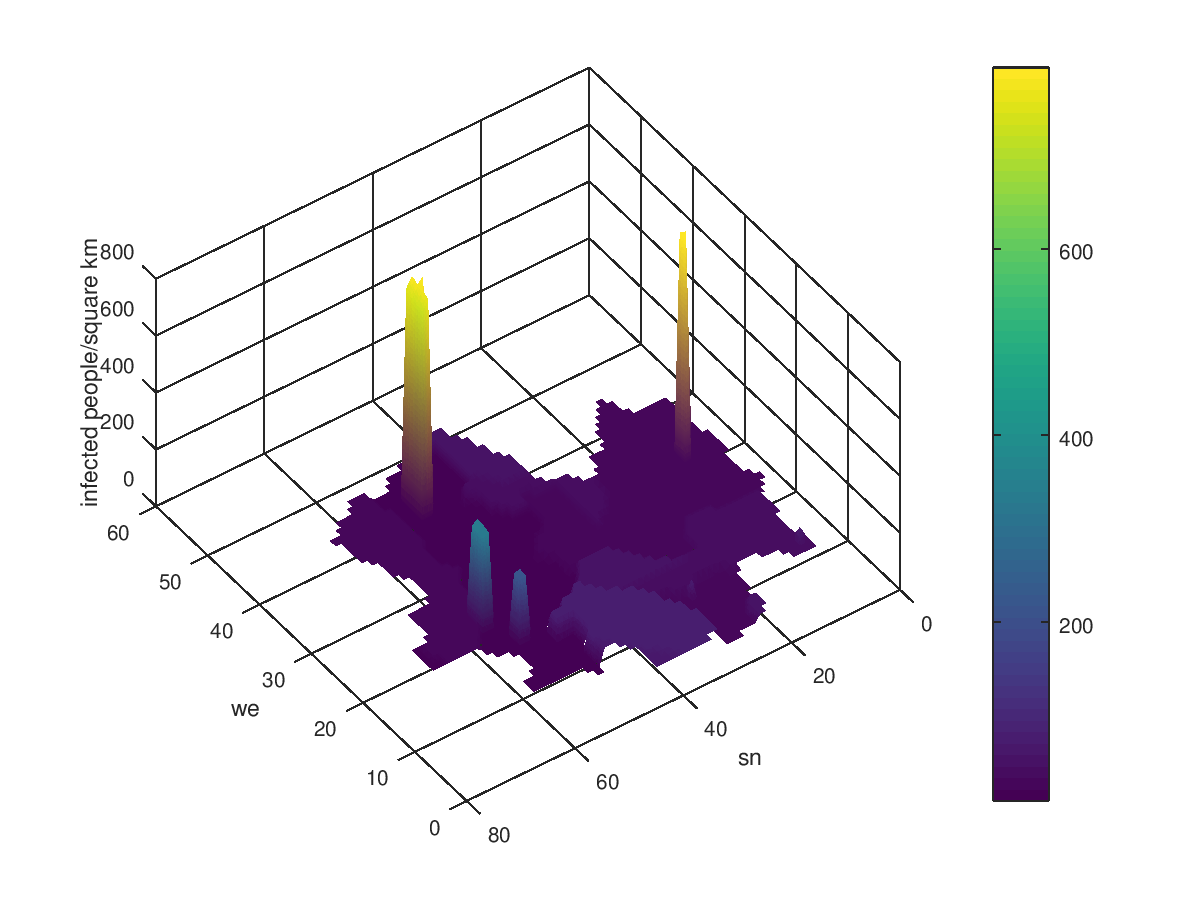}
                \caption{\label{fig52} Infected people after 200 time-steps $q=0$}
        \end{minipage}
\end{figure}

To get an idea of an appropriate approach of the source-sink term $q$ let us consider
the development of the 7-days incidence of three successive days in the federal states.
The guessed values of the change of $s$ per day are results of a linear regression.
The change of $s$ per day can be divede into a part coming from diffusion and another part
coming from the local passing on the virus. We assume a 10\% part coming from local transmission.
The figures \ref{fig61}-\ref{fig64} show the simulation results with the assumptions for $q$
after 10, 20, 30 and 50 time-steps (no border-crossing). The initial values are taken from 
table \ref{t1}, i.e. the $s$-data of 12.01.2021.

Because of the coarse approach of $q$ it is advisable to update $q$ based on the
actual data, especially if the local pandemic progression is changing stiffly.

\section{Discussion and Conclusion}
The examples show the impact of diffusion effects on the propagation
of the COVID-19 pandemic. It must be remarked that these processes are very
slow compared to the virus transmission in a local hotspot cluster. But with
the presented model is it possible to describe creeping processes which occur
beside slack measures like holey lockdowns.

Especially the pandemic propagation in regions with high incidence gradients can be described
with the discussed diffusion concept. 

It would be interesting to continue the work with diffusion models especially
the investigation of spreading events by a refined approach of
$q$ and the influence of border traffic. The presented model is qualified
for such investigations. But it's important to note that the diffusion 
modeling is only a small
part of the understanding of the expansion of the SARS-CoV-2 virus.

At the end it should be noted that the distribution and propagation must
be count back to the distribution and propagation of the
infected people. Doing this, one can see, that the dense peaks of infected
people are located in the metropolitan areas like Berlin, Munic or Hamburg. 
Figs. \ref{fig51}
and \ref{fig52} show the distribution of the seven-days incidence and
of the resulting infected people (by counting back using the people densities
and the area of the federal states).

This research received no external funding.

\section*{Acknowledgements}
Considering the topic of the pademic modeling I had some interesting exchange of ideas
with my friends and colleagues F. Bechstedt, physicist of the Friedrich-Schiller University Jena,
and Reinhold Schneider, mathematician of the Technical University Berlin. 
For this heartthanks.

\end{document}